\documentclass{amsart}
\usepackage{amsmath, amssymb, verbatim, amsthm, fullpage}

\usepackage[all]{xy}
\newcommand{\h}{\mathcal{H}}
\newcommand{\g}{\mathfrak{g}}
\newcommand{\R}{\mathbb{R}}
\newcommand{\C}{\mathbb{C}}
\newcommand{\Z}{\mathbb{Z}}
\newcommand{\dvol}{\textrm{ dvol}}

\newcommand{\Tr}{\textrm{Tr}}

\newcommand{\A}{\mathcal{A}}
\newcommand{\Gl}{{\rm Gl}}
\newcommand{\fr}{{\rm F}}
\newcommand{\bL}{{\mathbb L}}
\newcommand{\bg}{{\bar{g}}}
\newcommand{\bF}{{\bar{\phi}}}

\newcommand{\be}{\begin{eqnarray}}
\newcommand{\bes}{\begin{eqnarray*}}
\newcommand{\ee}{\end{eqnarray}}
\newcommand{\ees}{\end{eqnarray*}}
\newcommand{\Spec}{\textrm{Spec }}
\newcommand{\Div}{\textrm{div}}
\newcommand{\Hom}{\textrm{Hom}}
\newcommand{\id}{\operatorname{id}}

\newtheorem{theorem}{Theorem}[section]
\newtheorem{corollary}[theorem]{Corollary}
\newtheorem{lemma}[theorem]{Lemma}
\newtheorem{proposition}[theorem]{Proposition}

\theoremstyle{remark}
\newtheorem{remark}{Remark}

\theoremstyle{definition}
\newtheorem{definition}{Definition}

\begin{document}

\title{The geometric $\beta$-function in curved space-time under
  operator regularization} \author{Susama Agarwala}
\address{Oxford University, Mathmatical Institute, OX2 6GG}

\begin{abstract}
In this paper, I compare the generators of the renormalization group
flow, or the geometric $\beta$-functions, for dimensional
regularization and operator regularization. I then extend the analysis
to show that the geometric $\beta$-function for a scalar field theory
on a closed compact Riemannian manifold is defined on the entire
manifold. I then extend the analysis to find the generator of the
renormalization group flow to conformally coupled scalar-field theories on
the same manifolds. The geometric $\beta$-function in this case is not
defined.
\end{abstract}

\maketitle
\tableofcontents

\section{Introduction}
Quantum field theory (QFT) is a well understood phenomenon under the
assumption that the underlying space-time of this universe is flat. On a curved
background space-time, it is less well understood. One of the problems
is understanding renormalization and the renormalization scale
dependence of the quantum field theory. The $\beta$-function of a field
theory, which solves for the renormalization scale dependence of the
coupling constant, can be solved locally on a curved manifold, but not
globally \cite{We2}.

There is another, global, approach to studying the renormalization group flow for a given QFT \cite{HollandsWald03}. This paper takes a third approach. Connes and Marcolli developed a
renormalization bundle wherein one can write regularized
Feynman rules as sections of a principal bundle over a space
parameterized by the regularization parameter and renormalization
mass \cite{CMbook}. The renormalization group action defines a one parameter
diffeomorphism on this bundle. The regularized Lagrangian, as a
function of the renormalization mass parameter, corresponds to a one
parameter family of sections on this bundle. The vector space
generating this one parameter diffeomorphism is uniquely defined by a
specific flat connection on this bundle \cite{thesis1}.

For a special class of sections of this bundle, those with local
counterterms, the connection defining the renormalization group flow
generator is equisingular. The renormalization group flow in this case
is called the geometric $\beta$-function of the theory, named to evoke
the physical $\beta$-function, which is the solution to a differential
equation. The physical $\beta$-function is instrumental in solving the
regularized Lagrangian's dependence on the renormalization mass,
whereas the geometric $\beta$-function generates an associated curve in
the space of sections of the bundle. There is a surprising
relationship between Hopf algebras and a renormalization scheme
developed by Bogoliobov, Parasiuk, Hepp, and Zimmerman, (BPHZ
renormalization) that makes the geometric story possible.

In this paper, I show that operator regularization on flat space-time
defines a section with local counterterms in this renormalization
bundle. I calculate the geometric $\beta$-function for operator
regularization under BPHZ renormalization, similarly to the
calculations done by Connes and Marcolli for dimensional regularization
under BPHZ renormalization. Operator regularization has the advantage over dimensional
regularization in that it is well defined globally over a manifold. In particular, for a generic $n$ dimensional manifold, there is no canonical or clear way to extend this to an $n+z$ dimensional manifold, for $z$ a complex number. For a greater exposition and examples on this point, see \cite{Ha}. There have been other attempts to understand the action of the renormalization group on field theories over curved space time, such as \cite{HollandsWald03}. While the approach in this paper is not related to the Connes Kreimer approach to renormalization, it finds similar results to other papers in my program to understand the action of the renormalization group action on a variety of regularization methods, \cite{cutoffdimreg, dynkin}.

Unlike in \cite{HollandsWald03}, I only consider a closed compact Riemannian manifold $M$ with metric $g$. I extend the Connes Marcolli renormalization bundle to one that sits over this space and show
that operator regularization of a scalar QFT that has $M$ as a
background space can be written as a section of this new bundle, which
has local counterterms.  The renormalization group flow generator on
this section, its geometric $\beta$-function, generates a curve that
represents the renormalization mass dependence of a Lagrangian for a
QFT over such a background space-time. Specifically, I have described
the renormalization mass flow geometrically without resorting to local
solutions to differential equations.

Finally, I turn to scalar conformally coupled field theories, and adjust
operator regularization in this context to build a conformally
coupled operator regularized Laplacian that defines the free
theory. I extend the renormalization bundle to include this theory as
a section. Unfortunately, this section is not local, and therefore the
renormalization group flow generator is not equisingular, though it
can still be defined by a connection on the bundle.

In section \ref{tale}, I define operator regularization and dimensional
regularization for QFTs in flat space. I review the problem of scale
dependence and some general techniques of renormalization. I show that
the physical conditions imposed on well defined counterterms of a
regularized theory under renormalization implies locality for the
renormalization group action relevant to operator regularization and
dimensional regularization. Finally, I show that operator
regularization is a local section of the Connes Marcolli
renormalization bundle and derive its geometric $\beta$-function.

In section \ref{beyond}, I define global operator regularization over
the entire manifold $M$. I extend the renormalization bundle to this
setting, and show that the section representing operator
regularization globally is local. I then derive its geometric
$\beta$-function, and thus a geometric representation of the global
regularized Lagrangian's renormalization mass dependence.

In section \ref{conformal}, I examine scalar conformal field
theories. I first review the preliminaries of densities necessary for
the discussion, and then define the operator regularized conformally
invariant Lagrangian that defines the free conformal field
theory. I extend the renormalization bundle to this setting, and
derive the renormalization group flow generator for the section
corresponding to this regularization scheme under BPHZ
renormalization. Since this section is not local, the renormalization
group flow is not a geometric $\beta$-function.

\section{A tale of two regularizations\label{tale}}
In this section, I review operator regularization, and compare it to
dimensional regularization on $\R^n$. I also recall a few relevant
facts about renormalization and the Connes Marcolli renormalization
bundle. This material can be found either in standard physics text
books \cite{IZ} or in the existing literature, \cite{Collinsbook, El, CMbook}. Finally, I show that the
renormalization group flow for operator regularization can be
geometrically represented as a connection on the renormalization
bundle.

As a concrete example, consider Feynman integrals of a massive
$\phi^3$ theory in $\R^6$ \be \mathcal{L} = \frac{1}{2} \phi(-\Delta +
m^2)\phi + \lambda \phi(x)^3 \; .\label{phi3d6lagrangian}\ee Here the Laplacian and the Lagrangian density are understood to be defined over $\R^6$. Later in the paper, as space time varies, these will be labeled to denote the weak coupling with gravity.

\subsection{Operator and Dimensional Regularization \label{opdimreg}}
The definitions and exposition in this subsection can be found in many
standard physics textbooks. For instance, see \cite{IZ} for a primer
on Feynman diagrams, and \cite{Collinsbook} for a good exposition on
dimensional regularization. Operator regularization is covered in some
detail in \cite{El}.

\begin{definition}
A Feynman graph is an abstract representation of an interaction of
several fields. It is drawn as a connected, not necessarily planar,
graph with possibly differently labeled edges. The types of edges,
vertices, and the permitted valences are determined by the Lagrangian
density of the theory in the following way:

\begin{enumerate}
\item The edges of a Feynman graph are labeled by the different fields in
  the Lagrangian. For the Lagrangian in equation \ref{phi3d6lagrangian}, there is only one type of edge.
\item The composition of monomial summands with degree $>2$ in the
  Lagrangian density correspond to permissible valences and
  composition of internal vertices of the Feynman diagrams. The
  $\phi^3$ term in this Lagrangian means that all
  internal vertices have valence three.
\item Vertices of valence one are called external vertices. \end{enumerate}
\end{definition}

The building blocks of these Feynman graphs are called one particle
irreducible (1PI) diagrams.
\begin{definition}
A one particle irreducible graph is a connected Feynman graph
  such that the removal of any internal edge still results in a
  connected graph.
\end{definition}
All Feynman graphs associated to a theory can be constructed by
gluing together 1PI diagrams along exterior edges.

The Feynman rules associate an integral to each Feynman graph. All
calculations in this paper are done over a Riemannian manifold. Let $G(x,y)= \frac{1}{(2\pi)^6}\int_{\R^6}
\frac{e^{ip\cdot (x-y)}}{p^2 + m^2} d^6p$ be the Green's function for
the Laplacian on $\R^6$, \bes \Delta G(x,y) = \delta^{(6)}(x-y)
\;.\ees Then the Feynman rules in configuration space are:

\begin{enumerate}
\item If a graph $\Gamma$ has $I$ edges, write down the $I$ fold
  product of propagators, of various types according to the type of
  edges, \bes \prod_1^I G_{i}(x_i, y_i)\;, \ees where $x_i$ and $y_i \in \R^6$
  are the endpoints of each edge.
\item Each internal vertex, $v_i$, has valence $3$. Define a measure on
  $(\R^6)^{3}$ \be \mu_i = -i\lambda \delta(x_1, x_2)\delta ( x_2,
  x_3) \label{meas}\; ,\ee where the $x_i$ are the endpoints of the
  edges incident on the vertex in question in the graph, $\lambda$ is
  the coupling constant, and $\delta(x_i, x_j)$ is the Dirac delta
  function.
\item Integrate the product of propagators
  from above against this measure \be \int _{(\R^6)^{3V}}
  \prod_1^I G_{i}(x_i, y_i) \prod_i^V d\mu_i \; .\label{convolve}\ee
\item Divide by the symmetry factors of the graph.
\end{enumerate}

These rules can be generalized to other renormalizable theories such
as $\phi^4$ in $\R^4$. Conservation of momentum follows from taking the Fourier transform of these integrals. The Fourier transform
of these integrals gives a Feynman integral in momentum space of the form
\be \int_{\R^{6l}}\prod_{k=1}^I \frac{1}{f_k(p_i,e_j)^2 + m^2}
\prod_{i=1}^ld^6p_i \label{example}\; ,\ee where the $p_i$ are the
loop momenta assigned to each loop, $f(p_i,e_j)$ is a linear
combination of the loop and external momenta representing the momenta
assigned to each internal leg.

For a $\phi^3(x)$ scalar field theory, $n = 3$ for all $v_i$.
Since $\Delta^{-1}$ is not trace-class over $\R^6$,
the measure defined in equation \eqref{meas} forces the Feynman
integrals for most diagrams to be ill defined integral operators. In
order to make sense of the probability amplitudes they are supposed to
represent, the integrals need to be \emph{regularized}, or written in
terms of an extra parameter such that the new integrals are defined
away from a fixed limit. I consider two means of regularization
in this section, dimensional regularization and operator
regularization. The latter is also known as $\zeta$-function
regularization when restricted to one loop diagrams, or as analytic
regularization.

Write the integral in \eqref{example} in spherical coordinates, \bes
A(6)^l \int_0^\infty\prod_{k=1}^I \frac{1}{f_k(p_i,e_j)^2 + m^2}
\prod_{i=1}^lp_i^5 dp_i \; ,\ees where $A(d) =
\frac{\Gamma(d)}{(4\pi)^{d/2}}$ is the volume of $S^{d-1}$, the sphere
in $d-1$ dimensions. Dimensional regularization exploits the fact that
the integral above is convergent if taken over $d = 6+ z$, dimensions,
with $z$ a complex parameter. Notice that $A(d)$ is holomorphic in
$z$, and does not contribute to the polar structure of the graph. The
dimensionally regularized integral is \bes \varphi_{dr}(z)(\Gamma) =
A(d)^l \int_0^\infty\prod_{k=1}^I \frac{1}{f_k(p_i,e_j)^2 + m^2}
\prod_{i=1}^lp_i^{d-1} dp_i \; .\ees Put another way, dimensional
regularization assigns a holomorphic function, $A(d)$, times the
Mellin transform of each loop integral in the Feynman integral.  If
the original integral is divergent, this expression has a pole at $d =
6$.

For operator regularization, raise the Laplacian in the theory to a
complex power. The regulated Lagrangian density is, \bes
\mathcal{L}(x,z) = \frac{1}{2} \phi(x)(-\Delta + m^2)^{1+z}\phi(x) + \lambda
\phi(x)^3 \; ,\ees for $z \in \C$. The associated Green's function is
now \bes G^{1+z}(x,y)= \int_{\R^6} \frac{e^{ip\cdot (x-y)}}{(2\pi)^6(p^2 +
  m^2)^{1+z}} d^6p\;. \ees The Feynman integral in \eqref{example}
under operator regularization is \bes \varphi_{or}(z)(\Gamma) = A(6)^l
\int_0^\infty\prod_{k=1}^I \frac{1}{(f_k(p_i,e_j)^2 + m^2)^{1+z}}
\prod_{i=1}^lp_i^5 dp_i \; .\ees

\subsection{Scale dependence and renormalization}

Prior to regularization, the Lagrangian of any theory is dimensionless. The integral of the Lagrangian density over $\R^6$ does not
change under the coordinate change $x \rightarrow tx$ \be \int_{\R^6}
\mathcal{L}(x) \, d^6x = \int_{\R^6} \mathcal{L}(tx)\, d^6(tx)\;
.\label{reglag}\ee The regularized Lagrangian is not dimensionless. To make it so, it must be written as a function of both the regularization scale and the
regularization parameter, $\mathcal{L}(\mu, z)$.

This section deals with the problem of accounting for this extra
parameter in the Lagrangian, and how to extract finite quantities from
the regularized Lagrangian. Much of this material is found in standard
physics texts.  For further reading on the problem of renormalization,
see \cite{Collinsbook}.

I begin with some standard dimensional analysis. Using the
convention where $c = \hbar = 1$, the mass and energy dimension of a
function are the negative of the length dimension of the
function.

\begin{definition}
Denote by $[P(x)]$, the degree of the function $P$. That is, \bes P(tx) = t^{[P(x)]} P(x) \;.\ees  \end{definition}

Since, in configuration space, $x$ represents distance, $[P(x)]$ represents the length dimension of $P(x)$.


The component terms of the unregularized Lagrangian have the following
dimensions: \be \begin{array}{cc} \left[\phi(x)\right] &= -2
  \\ \left[\Delta\right] &= -2 \\ \left[ m \right] & = -1\\ \left[\lambda\right] &= 0
  \\ \left[\dvol \right] &= 6 \; .\end{array} \label{dims}\ee The
conformal dimension of the Laplacian raised to a power is \bes
\left[(-\Delta +m^2)^{1+z}\right] = -2(1+z)\; .\ees Similarly, the dimension
of the volume form $\R^{6+z}$ is \bes
\left[\dvol(\R^{6+z})\right] = (6+z).\ees The unregularized
Lagrangian is dimensionless, the dimension of the Lagrangian density
is $\left[\mathcal{L}(x)\right]=-6$.

The operator regularized Lagrangian is \bes L(
z)=\int_{\R^6} \phi(x)(-\Delta+m^2)^{1+z}\phi(x) + \lambda \phi^3(x) d^6 x
\;. \ees This is neither homogeneous nor dimensionless unless
$[\lambda]= -2z$.  This can be done by introducing a renormalization
mass factor to the coupling constant $\lambda \rightarrow \lambda\mu^{-2z}$ where $[\mu] =
-1$. The dimensionless regularized Lagrangian is \be L(z) =
\int_{\R^6} \left[\frac{1}{2} \phi(x)(-\Delta +m^2)^{1+z} \phi(x)
  +\lambda\mu^{2z} \phi(x)^3 \right] \mu^{-2z} d^6x \label{scalingor}\;.\ee Similarly, the scale dependent Lagrangian under
dimensional regularization is \be L(z)
= \int_M \left[\frac{1}{2} \phi(x)(-\Delta +m^2) \phi(x)
  +\lambda \phi(x)^3 \right] \mu^{-z}\dvol(x) \; .\label{scalingdr}\ee Here, $\mu$ is the energy scale, or the T'hooft mass of the theory. Varying $\mu \in \R_+$ is the action of the renormalization scale on $L$.

To get finite quantities out of a regularized theory, the Lagrangian
is renormalized. The coefficients and fields defining regularized
Lagrangians are called bare quantities. Following \cite{Collinsbook}, I
write the regularized Lagrangian density over a manifold $M$ \bes \mathcal{L}_M(x,z) =
(\partial \phi_0(x))^2 + m_0^2\phi^2_0(x) + \lambda_0\mu^{2z}
\phi_0^3(x) \;. \ees I then rescale the bare field by a function of
$\mu$, $\phi_0(x) = Z_\phi^{1/2} (\mu) \phi(x)$, where $\phi(x)$ is
the renormalized field when $Z_\phi=1$.  In terms of these
renormalized quantities, the bare Lagrangian is \bes \mathcal{L}_M(x,z) =
Z_\phi (\partial \phi(x))^2 + m_0^2Z_\phi\phi^2(x) +
\lambda_0Z_\phi^{3/2}\mu^{2z} \phi^3(x) \;. \ees Write the bare mass
$m_b = m_0Z_\phi^{1/2}$ and the bare coupling constant $\lambda_b
=\lambda_0Z_\phi^{3/2}$. These can both be written in terms of the regularized quantities, $m^2_b = Z_mm^2$ and
$\lambda_b = Z_\lambda \lambda$ respectively.  In this notation, the Lagrangian density
can be split into a finite and counterterm part \be \mathcal{L}_M(x,z) &=&
(\partial \phi(x))^2 + m^2\phi^2(x) + \lambda\mu^{2z}
\phi^3(x) \label{fp}\\ &+& (Z -1)(\partial \phi(x))^2 +
(Z_m-1)m^2\phi^2(x) + (Z_\lambda -1)\lambda\mu^{2z} \phi^3(x)
\;. \label{ct}\ee The quantities $\sqrt{(Z-1)}\phi(x)$, $(Z_m-1)m^2$
and $(Z_\lambda -1)\lambda\mu^{2z}$ are the counterterms of the
theory, which cancel the divergences of the theory in the limit $z
\rightarrow 0$. The expression in \eqref{fp} is called the finite
Lagrangian density, and \eqref{ct} the counterterm Lagrangian density,
\bes \mathcal{L}_M(x, z) = \mathcal{L}_{M,fp} + \mathcal{L}_{M,ct}\;.\ees The
quantities $\phi_0(\mu, z, x)$, $m_0(\mu, z)$ and $\lambda_0(\mu, z)$
all depend on the scale of the theory, $\mu$, and the regularization
method. A set of differential equations, called the renormalization
group equations, solve for the scale dependence of the various quantities in the regularized Lagrangian. The $\beta$-function of a
Lagrangian under a specific regularization scheme is the dependence of
the coupling constant on the renormalization scale \bes
\beta(\lambda_0) = \mu \frac{\partial}{\partial \mu} \lambda_0 \;
.\ees  It is
calculated perturbatively by loop number of the Feynman diagrams.

One can use different renormalization algorithms to calculate the
counterterms of individual Feynman diagrams.

\begin{definition}
Let $U(\Gamma)$ be the unrenormalized Feynman integral. For a given
renormalization scheme and regularization scheme, let $C(\Gamma)$ be
the counterterm defined, and $R(\Gamma) = U(\Gamma) - C(\Gamma)$ the
renormalized quantity.
\end{definition}

For a renormalization scheme and regularization scheme pair to be
physically significant, one needs to impose some conditions on the
counterterms.

\begin{enumerate}
\item \label{homo}The counterterms of a Feynman graph $\Gamma$ are homogeneous
  functions of the mass, and external momenta of degree
  less than or equal to the superficial degree of divergence of the
  graph, $6 -2 E(\Gamma)$, where $E(\Gamma)$ is the number of external
  legs of the graph.
\item \label{finite}The renormalized quantity is finite.
\end{enumerate}

Specifically, a counterterm is not well defined if its
\emph{dimension} varies with the regularization parameter.

\begin{definition}
If $C(\Gamma)$ satisfies condition \ref{homo} above, and $R(\Gamma)$
satisfies \ref{finite}, under regularization scheme $A$ and
renormalization scheme $B$, then I write that $A$ has well defined
counterterms under $B$.
\end{definition}

For instance, dimensional regularization has well defined counterterms
under BPHZ renormalization \cite{Collinsbook, CK00}. Speer shows that
operator regularization has well defined counterterms under BPHZ
renormalization \cite{Speer68, Speer74}. The factor of $\mu^{2z}$ and $\mu^z$ in the Lagrangians for operator \eqref{scalingor} and dimensional \eqref{scalingdr} regularization translate to a multiplicative factor of
$\mu^{zl(\Gamma)}$ in the Feynman integral of the graph $\Gamma$,
where $l(\Gamma)$ is the loop number of $\Gamma$.

Geometrically, for operator and dimensionally regularized theories,
$U(\Gamma) \in \Hom_{lin}(C^\infty(\R^{6l(\Gamma)}), \C\{\{z\}\})$ is
a linear map from the external momentum data to Laurent polynomials
with poles of finite degree. Varying the renormalization mass in
$U(\Gamma)$ gives a one parameter path in
$\Hom_{lin}(C^\infty(\R^{6l(\Gamma)}), \C\{\{z\}\})$. The renormalized
Feynman rules are a linear map from the vector space generated by the
1PI graphs to $\Hom_{lin}(C^\infty(\R^{6l(\Gamma)}), \C\{\{z\}\})$. In
the next section I impose a Hopf algebraic structure on the 1PI
diagrams to study the action of the renormalization mass scale.

\subsection{The renormalization bundle}

In \cite{CK00}, Connes and Kreimer build a Hopf algebra, $\h$, out of
the divergence structure of the Feynman diagrams for a scalar field
theory under dimensional regularization.  The co-product of the Hopf
algebra is defined to express the same sub-divergence data as in
Zimmermann's subtraction formula for BPHZ renormalization \cite{IZ}.

To briefly recall notation, let \bes \h = \C[\{1PI \text{ graphs with
    2 or 3 external edges, internal valence 2 or 3}\}] \ees be the
Hopf algebra of Feynman diagrams, with multiplication defined by
disjoint union. It is graded by loop number, with $Y$ the grading
operator. If $\Gamma \in \h_n$, $Y(\Gamma) = n\Gamma$. The co-unit
$\epsilon$ is $0$ on $\h_{\geq 1}$, and is the identity map on
$\h_0$. An admissible sub-graph of a 1PI graph, $\Gamma$, is a graph,
$\gamma$, or product of graphs, that can be embedded into $\Gamma$
such that each connected component has 2 or 3 external edges. The
graph $\Gamma/\gamma$ is that obtained by replacing each
connected component of $\gamma$ with a vertex. The admissible
sub-graphs correspond to the divergences subtracted by Zimmermann's
subtraction algorithm. Using Sweedler notation, the co-product on $\h$
is given by the sum \bes \Delta \Gamma = 1 \otimes \Gamma + \Gamma
\otimes 1 +\sum_{\gamma \text{ admis}}\gamma \otimes \Gamma//\gamma \;
.\ees Here, the symbol $\Gamma//\gamma$ refers to the contraction of $\Gamma$ by the connected components of $\gamma$. Let $\epsilon$ and $\eta$ denote the co-unit and unit of this
Hopf algebra.

The Hopf algebra is connected and each graded component $\h_n$ is
finitely generated as an algebra. Write the graded dual of this Hopf
algebra $\h^* = \oplus_n \h_n^*$. The product on $\h^*$ is the
convolution product $f\star g (\Gamma) = m(f \otimes g)
\Delta(\Gamma)$. The antipode, $S$, on the restricted dual defines the
inverse of a map under this convolution product, $f^{\star -1} = S
(f)$. By the Milnor-Moore theorem, $\h^* \simeq \mathcal{U}(\g)$ is
isomorphic to the universal enveloping algebra of the Lie algebra
$\g$, generated by the infinitesimal derivatives \bes
\delta_\Gamma(\Gamma') = \begin{cases} 1 & \Gamma = \Gamma' \text{
    1PI} \\ 0 & \Gamma \neq \Gamma' \end{cases} \;.\ees The generators
of the Lie algebra are infinitesimal characters \bes
\delta_\Gamma(\gamma\Gamma') = \epsilon(\gamma) \delta_\Gamma(\Gamma')
+ \epsilon(\Gamma') \delta_\Gamma(\gamma) \;.\ees

Both operator regularization and dimensional regularization evaluate
to $\A = \C\{\{z\}\}$, formal Laurent polynomials in $z$. Let $\pi :
\A \rightarrow z^{-1}\C[z^{-1}] = \A_-$ be the projection operator
onto the polar part of the Laurent series. Define $\A_+ =
\C[[z]]$. The algebra $\A = \A_+ \oplus \A_-$.

\begin{definition} Write the Feynman
integral of the Feynman diagram $\Gamma$ under operator regularization
and dimensional regularization as $\varphi_{or}(\Gamma)$ and
$\varphi_{dr}(\Gamma)$. Then $\varphi_{or} , \varphi_{dr} \in
\Hom_{alg}(\h, \A)$ is the algebra homomorphisms from $\h$, the Hopf
algebra of Feynman graphs, to $\A$, the algebra spanned by the
regulating parameters. \end{definition}

In \cite{CK00}, Connes and Kreimer show that BPHZ renormalization can
be written as a composition of loops in the Lie group $G$ using the
Birkhoff decomposition theorem. Here, the Lie group $G$ is the affine group scheme associated to the Hopf algebra $\h$, and the Lie group associated to the Lie algebra defining $\h^\vee$: \bes G = \textrm{exp } \g \;. \ees The following is a summary of their
results.

Consider the punctured infinitesimal disk around the origin in $\C$, written $\Delta^*
=\Spec \A$. Let $\gamma(z)$ be a map from a simple loop not containing
the origin in $\Delta^*$ to $G$. If $K$ is a trivial principal $G$
bundle over the space $\Delta^*$, then the maps $\varphi(z)$ are
sections of this bundle. There is a natural isomorphism from the group
of these maps and $G(\A)$.  By the Birkhoff decomposition theorem,
$\varphi(z)$ decomposes as the product \bes \varphi(z) =
\varphi(z)_-^{\star -1}\star \varphi(z)_+ \;,\ees where $\varphi_+(z)$ is a
well defined map in the interior of the loop (containing $z=0$), and
$\varphi_-^{\star -1}(z)$ is a well defined map outside of the loop (away
from $z=0$). Each $\varphi(z)$ can be written as a Laurent series with
poles of finite order and coefficients in $G(\C)$ convergent in
$\Delta^*$.  The map $\varphi(z)_+$ is a somewhere convergent formal
power series in $z$, and for $x_\Gamma \not \in \ker(\varepsilon)$,
$\varphi(z)_-(x_\Gamma) = \sum_{-n}^{-1} a_i z^i$, where $a_i \in
G(\C)$. Finally, normalizing $\varphi(z)_-(x_\emptyset) = 1_\h$,
ensures the uniqueness of the Birkhoff decomposition. Explicitly, \bes
\varphi(z)_-(\Gamma) = -\pi(\varphi(z)(\Gamma) + \sum_{\gamma \text{
    admis.}}\varphi(z)_-(\gamma)\varphi(z) (\Gamma//\gamma)
\\ \varphi(z)_+(\Gamma) = (\id-\pi)(\varphi(z)(\Gamma) + \sum_{\gamma \text{
    admis.}}\varphi(z)_-(\gamma)\varphi(z)(\Gamma//\gamma)) \;. \ees

Connes and Kreimer, in loc. cit., show that the recursive formula for
calculating $\varphi(z)_+(x_\Gamma)$ and $\varphi(z)_-(x_\Gamma)$ is
exactly the same as the recursive formula for calculating the
renormalized and counterterm contributions respectively of a Feynman
diagram $\Gamma$ to the regularized Lagrangian given by BPHZ. For
$\Gamma$ a 1PI graph, $\varphi(z)(x_\Gamma)$ is the value of the
regulated Feynman integral of the graph $\Gamma$, $\lim_{z\rightarrow
  0}\varphi_+(z)(x_\Gamma)$ is the renormalized value of the graph
while $\varphi_-(z)(x_\Gamma)$ is the counterterm.

In the geometric presentation of the renormalization, generalize the
renormalization group to $\C^\times$ \cite{CMbook}. Then the
renormalization group action is \bes \C^\times \times G(\A)
&\rightarrow& G(\A) \\ (t, \varphi) &\rightarrow& t^Y \varphi \;.\ees
Incorporating the renormalization group into the regularization bundle
gives a new bundle $P \rightarrow B$ with $P \simeq K \times
\C^\times$, $B = \Delta^*\times \C^\times$.

\begin{definition} Let $\tilde{G}(\A) = G(\A) \rtimes
\C^\times$ be the group with multiplication \bes (\varphi(z), t) \star
(\psi(z), u) = (\varphi(z)\star t^Y \psi(z), tu) \ees with
$\varphi(z), \psi(z) \in G(\A)$ and $t, u \in \C^\times$. Let
$\tilde\g(\A)$ be the Lie Algebra of $\tilde{G}(\A)$.  \end{definition}

As constructed, $P$ is a trivial $\tilde{G}(\A)$ principal over
$B$. It is a $\C^*$ equivariant bundle \bes t \circ (\varphi(z), u)
\rightarrow (t^Y\varphi(z), tu) \;.\ees

In \cite{thesis1}, I show that there is a global flat connection,
$\omega(z,t) \in \Omega^1(\tilde\g(\A))$, on $P$ defined by the
bijection \bes \tilde{R} : G(\A) &\rightarrow& \g(\A) \\ \varphi(z)
&\rightarrow& \varphi^{\star -1}(z) \star Y \varphi(z) \;.\ees The
properties of the $\tilde{R}$ bijection are discussed in \cite{EM}. By
$\C^\times$ invariance of $P \rightarrow B$, it is sufficient to study
$\omega(z,t)$ pulled back along the sections of the form
$(t^Y\varphi(z), 1)$.  These pullbacks, $(t^Y\varphi)^*\omega(z, t)$
define connections on $B$ and can be written \bes
(t^Y\varphi)^*\omega(z,t) = t^Y\varphi^{\star -1}(z) \star d
t^Y\varphi(z)\;.\ees More generally, a connection on $B$ is of the
form \bes \omega' = a dz + b dt \, \ees with $a, b \in \g(\A)$. Define
$\psi = \tilde{R}^{-1}(\frac{b}{z}) \in G(\A)$. If $a = \psi^{\star
  -1} \star \frac{\partial}{\partial z} \psi$ then $\omega' =
\psi^*\omega(z,t)$.

The vector field $\tilde{R}(\varphi(z))$ is the generator of the
renormalization group flow of $\varphi(z)$ under the renormalization
group action $t \circ \varphi(z) = t^Y \varphi(z)$. This action is
appropriate for only certain regularization schemes, including dimensional and operator regularization. Other
regularization schemes have different group actions.

\begin{definition}
Let $\varphi(z)$ be a section of $K\rightarrow \Delta^*$. It is local
if \bes \frac{\partial}{\partial t} (t^{Y}\varphi(z))_- = 0 \;.\ees
\end{definition}

The following theorem is a generalization of results in \cite{CM06}, where the authors only consider dimensional regularization. While the result is similar to that shown by the authors of loc. cit., the proof is different.

\begin{theorem}
Let $\varphi(z)$ be a section of $K\rightarrow \Delta^*$ that
represents a regularized QFT with renormalization group action \bes t
\circ \varphi(z)(\Gamma) = t^{zY(\Gamma)}\varphi(z)(\Gamma) \;,\ees
that has well defined counterterms under BPHZ, then $\varphi(z)$ has
local counterterms. \label{localthm}
\end{theorem}

\begin{proof}
The renormalization group action gives
$t^{zY(\Gamma)}\varphi(z)(\Gamma)$ which Birkhoff decomposes as \bes
t^{zY}\varphi(z)_-^{\star -1} \star t^{zY}\varphi(z)_+ (\Gamma) \ees since
$t^{zY}$ is an automorphism on $G(\A)$.

Write the counterterm of the Feynman diagram as $\varphi(z)_-(\Gamma)
= C(p_1 \ldots p_{E(\Gamma)}, m^2, z)(\Gamma)$, for $p_i$ the external
momenta of the Feynman integral of $\Gamma$. On the level of Feynman
integrals, the renormalization group action is given by the change of
loop momenta variable $p_i \rightarrow tp_i$, so the new counterterm
is \bes t^{-k} C(tp_1 \ldots tp_{E(\Gamma)}, tm, z)(\Gamma)\ees for some
integer $k$ that depends only on $\Gamma$. However, the
renormalization group action also means \bes t^{-k} C(tp_1 \ldots
tp_{E(\Gamma)}, tm, z)(\Gamma) = t^{zY(\Gamma)}C(p_1, \ldots,
p_{E(\Gamma)}, m, z)(\Gamma) \;.\ees Since $k$ is constant, the right
hand side cannot depend on $z$. In other words, \bes \frac{d}{dt}
t^{zY(\Gamma)}C(p_1, \ldots, p_{E(\Gamma)}, m, z)(\Gamma) = 0 \ees or
\bes \frac{\partial}{\partial t} (t^{Y}\varphi(z))_- = 0 \;, \ees as
desired.
\end{proof}

\begin{definition}
The geometric $\beta$-function of a section is defined \bes
\beta(\varphi(z)) = \lim_{z\rightarrow 0}(t^{Y}\varphi(z))^{\star -1}
\star t \frac{\partial}{\partial t} (t^{Y}\varphi(z))|_{t=1}\;.\ees
\end{definition}

Ebrahimi-Fard and Manchon show that $\beta(\varphi(z)) =
z\tilde{R}(\varphi(z)_-)$ if $\varphi(z)$ has local
counterterms \cite{EM}.

\begin{corollary}
The geometric $\beta$-function for $\varphi_{or}(z)$ and
$\varphi_{dr}(z)$ is well defined.
\end{corollary}
\begin{proof}
Since dimensional regularization and operator regularization both have
well defined counterterms under BPHZ renormalization, the corresponding
sections of $K\rightarrow \Delta^*$, $\varphi_{dr}(z)$ and
$\varphi_{or}(z)$ are local. \end{proof}

It is important here to clarify a distinction between the physical
$\beta$-function of a theory and the geometric $\beta$-function first
defined by Connes and Kreimer in \cite{CK01}. The renormalization
group action $t^Y\varphi(z)$ defines a one parameter diffeomorphism of
$G(\A)$, the renormalization group flow. The vector field
$\tilde{R}(\varphi(z))$ defines the diffeomorphism. On sections with
local counterterms, it restricts to the $\beta(\varphi)$.

The physical $\beta$-function, on the other hand, is the dependence of
the coupling constant on the renormalization mass. It is used to
calculate the dependence of the regularized Lagrangian on the
renormalization mass. This is not defined if the regularization scheme
does not have well defined counterterms under a renormalization
scheme.

\section{Beyond $\R^n$ \label{beyond}}

In this section I extend the analysis over $\R^n$ to a scalar field
theory over a compact closed manifold $M$. I then define the
renormalization group flow that is well defined over the entire
manifold $M$.

First I develop the Feynman rules in configuration space for a
renormalizable scalar field theory with valence three interactions on
a Riemannian manifold, $(M, g)$, with metric tensor $g$. I want to
develop a theory that is coherent globally over $M$, though
calculations must be done locally. Therefore, I ignore the issue of
Wick rotation from a Lorentzian metric, since Wick rotation is a local
map without a global counterpart. I also cannot work in momentum space,
as translating from configuration to momentum space requires Fourier
transforms, a local operation.

Let $(M,g)$ be a $6$ dimensional Riemannian manifold with metric $g$. Write the Lagrangian densities on this manifold as \be
\mathcal{L}_M= \frac{1}{2} \phi(-\Delta(g) +m^2)\phi + \lambda\phi^3 \;
, \label{StLag} \ee where $m$ is the mass parameter, and $\lambda$ is
the coupling constant and $\Delta(g)$ the Laplacian on the manifold \bes
\Delta(g) = \Div\circ\nabla = \frac{1}{\sqrt|g|}
\partial_i(\sqrt{|g|}g^{ij}\partial_j)\; .\ees To simplify notation, I call the mass shifted Laplacian on $M$, $\Delta_M = -\Delta(g)
+m^2$. Then \bes
\mathcal{L}_M= \frac{1}{2} \phi(\Delta_M)\phi + \lambda\phi^3 \;
, \label{StLag} \ees
The Lagrangian on the manifold is then given by \bes L_M =
\int_M \mathcal{L}_M \sqrt{|g(x)|}d^nx \; ,\ees where $|g| = \det
g_{ij}$. More details can be found in standard physics texts
such as \cite{IZ}.

\subsection{Regularization over $M$ \label{regcomp}}

Working in configuration space, a Feynman integral is an integral
operator that can be written as a \emph{generalized} convolution
product of Green's functions associated to the Laplacian on the
manifold $\Delta_M$. This integral operator is not well defined for
$\dim (M) > 1$, since $\Tr (\Delta_M)^{-1}$ is not well defined in
this case. These integral operators require regularization. Dimensional
regularization is not well defined globally over a general manifold. There is no canonical way to extend a $n$ dimensional manifold to a non-integral, complex, $n+z$ dimensional manifold. On the other hand, operator regularization, which works by raising the Laplacian to a complex
power, is well defined on general manifolds \cite{Ha}. In this section, I consider operator regularized Feynman
integrals in configuration space over a closed compact Riemannian
manifold.

The regularized Feynman integral acts on the symmetric algebra $S(E)=
\oplus_n S^n(E)$ over the external vertex data, $E =
C^\infty(M)$. Each external vertex of a Feynman diagram is assigned a
function $f \in E$. If a Feynman diagram has $n$ external edges, it
acts on $S^n(E)$.

The Feynman rules in this setting are the same as those defined in
section \ref{opdimreg} except that the flat space Green's kernel is
replaced by the Green's kernel associated to $\Delta_M$. This is a
distribution on $(M\times M)$, $G_M(x,y) \in \mathcal{D}'(M\times M)$,
defined by the equation \bes \Delta_M G_M(x,y) = \delta(x,y)\; ,\ees
where $\delta(x,y)$ is the Dirac delta function. Notice that due to
the structure of $\Delta$, the Green's function $G_M$ depends on $g$,
the metric on $M$. Since $-\Delta_M+m^2$ is a negative (semi)-definite
elliptic operator acting on $E = C^\infty(M)$, for a compact manifold
$M$, it has a discrete spectrum that can be ordered \bes 0 \geq
\lambda_1 \geq \ldots \lambda_i \geq \lambda_{i+1} \ldots\ees
including multiplicity, and an orthonormal set of eigenfunctions
$\{\phi_i\}$ such that \bes \int_M \phi_i(x) \phi_j(x) \dvol(x) =
\delta_{ij}\;,\ees where $\delta_{ij}$ is the Kronecker delta
function. Write $E = E_0 \oplus E_-$, where $E_0 = \ker(\Delta_M)$ and
$E_- = \oplus_iE_i$ the direct sum of the negative eigenspaces of
$\Delta_M$. The Green's function $G_M(x,y)$ is the inverse of
$\Delta_M|_{E_-}$ and can be written \bes G_M(x,y) = \sum_{i=1}^\infty
\frac{\phi_i(x) \phi_i(y)}{\lambda_i}\; .\ees

Seeley shows that for a 6-dimensional manifold, $M$, $\Delta_M^{1+z}$
is trace class, with $G_M^{1+z}(x,x)$ meromorphic with simple poles at
$1+z = k-3$, where $k \in \mathbb{Z}_{\geq 0}$ \cite{Se}. The
corresponding Green's function is \bes G_M^{1+z}(x,y) = \sum_{i =
  1}^\infty \frac{\phi_i(x)\phi_i(y)}{\lambda_i^{1+z}}\;,\ees with
\bes \Tr(\Delta_M^{-(1+z)}) = \int_M G_M^{1+z}(x,x)\dvol(x) =
\sum_{i=1}^\infty \frac{1}{\lambda_i^{1+z}}\; .\ees The quantity
$\sum_{i=1}^\infty \frac{1}{\lambda_i^{1+z}} = \zeta (\Delta_M)$ is
called the operator $\zeta$-function for the Laplacian $\Delta_M$.
Since $G_M^{1+z}(x,x)$ is meromorphic in $z$, $\sum_{i=1}^\infty
|\frac{1}{\lambda_i^{1+z}}| < \infty $ and $\Delta_M^{1+z}$ is trace
class for $z \not \in \Z$.

\begin{proposition} The regularized Feynman rules for a graph $\Gamma$
is a Schwartz kernel $K^{reg}_\Gamma \in \mathcal{D}' (M^{|E(\Gamma)|})
\{\{z\}\}$ that can be written as a somewhere convergent Laurent
polynomial with finite poles at $0$ and distribution valued
coefficients. \label{distregfr}\end{proposition}

\begin{proof}
First consider the heat operator, $e^{t\Delta_M}$, with $t \in
\R_{>0}$. It is related to the complex powers of the Laplacian, $\Delta_M$
by a Mellin transform: \be \Gamma(z) (-\Delta_M)^{-z-1}f(x) =
\int_0^\infty \exp(t \Delta_M)f(x) t^z \; dt \; , \label{cmpxdelta}
\ee where $f(x) \in C^\infty(M)$, and $z \in \C$. The heat operator
has a unique kernel $G_M(t, x, y)$, which is continuous on $M \times
M$, and smooth away from the diagonal. The Schwartz Kernel theorem
associates a kernel to $\Delta_M^{-1-z}$, $G^{1+z}_M(x,y)$. Equation
\eqref{cmpxdelta} shows that $G_M^{1+z}(x,y)$, \be G_M^{1+z}(x,y) =
\frac{-1^{-1-z}}{\Gamma(z)}\int_0^\infty t^zG_M(t, x, y) \; dt \;
. \label{cmpxgreen} \ee To see $G^{1+z}_M$ more explicitly, let
$\Re(z) >\dim M$. The right hand side of \eqref{cmpxgreen} is well
defined in this region. Then $G^{1+z}_M$ can be defined by
analytically continuing to all $z$.

The regularized Feynman integrals now look like \bes
K^{reg}_\Gamma(x_1, \ldots x_{|E(\Gamma)|}) = \int_{M^{|V(\Gamma)|}}
\prod_1^I G_M^{1+z} (x_{i_ 1}, x_{i_2}) \dvol (x_1,\ldots x_V) \ees
where $E(\Gamma)$ is the set of external vertices of $\Gamma$, $V(\Gamma)$
the set of internal vertices, $I(\Gamma)$ the set of internal edges and $i_1
\,,i_2 \in \{1 \ldots |V(\Gamma) \cup E(\Gamma)|\}$. Substituting in \eqref{cmpxgreen}
gives \bes K^{reg}_\Gamma(x_1, \ldots x_{|E(\Gamma)|}) =
\\ (\frac{(-1)^{-1-z}}{\Gamma(z)})^{|I(\Gamma)|}\int_0^\infty \int_{M^{|V(\Gamma)|}}
\prod_1^I G_M(t_i, x_{i_1}, x_{i_2}) \; \dvol (x_1,\ldots x_V)
\prod_{i=1}^It_i^z dt_i \; .\ees

\end{proof}

\begin{corollary}
For a Feynman diagram $\Gamma$ with $n$ external legs, there is an
operator, $A_\Gamma(z)$, associated to $K^{reg}_\Gamma(z)$ by the
Schwartz kernel theorem which can be written as \bes
A_\Gamma(z):S^{n}(E) \rightarrow \C\{\{z\}\}\;, \ees a linear map from
the external leg data to the space of somewhere convergent Laurent
polynomials with finite order poles at $0$ and $\C$
coefficients. \label{opregfr} \end{corollary}

\begin{proof}
This follows from the trace properties of $\Delta^{1+z}_M$ \end{proof}
This regularization method is sometimes called analytic
renormalization, as in \cite{Speer68}. If computations are done in
local coordinates, and the Feynman integrals are studied in momentum
space, this is called operator regularization \cite{El}.

By Proposition \ref{distregfr}, one can view the regularized Feynman
rules as maps from Feynman diagrams to $\mathcal{D}'(M^{|E(\Gamma)|})\{\{z\}\}$. This is the view taken in the physics literature,
where Feynman integrals are referred to as Green's functions. In this
paper, I am also interested in the corresponding integral operator
given by Corollary \ref{opregfr}.

\begin{theorem}
Let $\Gamma$ be a Feynman diagram, with $|E(\Gamma)|$ external vertices, and $A_\Gamma(z)$ the operator associated to it by Corollary \ref{opregfr}. Then, for $f \in S^i(E)$,
and $h \in S^j(E)$ where $i + j = |E(\Gamma)|$, $\langle A_\Gamma(z) f,
h\rangle$ depends only on the metric $g$ of the base manifold $M$, and
the combinatorics of the graph $\Gamma$. \label{resg}
\end{theorem}

\begin{proof}
As before, let $V(\Gamma)$ be the set of internal vertices of $\Gamma$ and $I(\Gamma)$ the set
of internal edges. It is sufficient to regularize $\langle A_\Gamma(z) f,
h\rangle$ where the external edge data $f =
\prod_{k=1}^i\phi_{f_k}(x_{f_k})$, $h =
\prod_{k=1}^{j}\phi_{h_k}(x_{h_k})$ are products of eigenfunctions of
$\Delta_M$, each associated to the external vertex $x_{f_k}$ or
$x_{h_k}$.  Write \bes \langle A_\Gamma(z) f, h \rangle =
\int_{M^{|V(\Gamma)|}}fh\prod_{i=1}^{|I(\Gamma)|} \sum_{k_i=0}^\infty\frac{\phi_{k_i}
(x_{i_1})\phi_{k_i}(x_{i_2})}{\lambda_{k_i}} \dvol(x_1, \ldots,
x_{|V(\Gamma)|})\ees with $x_{i_1}, \, x_{i_2}\in \{x_1, \ldots, x_{|V(\Gamma)|}\}$. This can
be regularized using the Mellin transform \bes \frac{1}{\lambda^{1+z}} =
\frac{1}{\Gamma(z)}\int_0^\infty t^{z} e^{-t\lambda}dt\ees then \bes
\langle A_\Gamma(z)f, h \rangle = \frac{1}{\Gamma(1+z)^{|I(\Gamma)|}}
\int_{M^{|V(\Gamma)|}}fh \sum_{k_1 \ldots k_I = 0}^\infty
\phi_{k_i}(x_{i_1})\phi_{k_i}(x_{i_2}) \\ \int_0^\infty\prod_{i=1}^{|I(\Gamma)|}
e^{-t_i\lambda_{k_i}}t_i^z dt_i \dvol(x_1, \ldots, x_{|V(\Gamma)|})\; .\ees
Integrating against the vertex variables gives \bes \int \sum_{i,j,k} \phi_i(y) \phi_j(y) \phi_k(y)dy = \int
\sum_{i,j,k} \phi_i(y) \sum_l a^{jk}_l \phi_l(y)dy = \sum_{j,k,i}
a^{jk}_i\; ,\ees where $\phi_j(y) \phi_k(y) =
\sum_la^{jk}_l\phi_l(y)$.  Since the quantity $a_i^{jk}$ is symmetric
on $i,\, j$, and $k$ I write it instead as $a(\{i, j, k\})$. The
quantity $a(\{i,\,j,\,k\})$ is tensorial, and depends only on the
metric of $M$. Define a function $f_v = \{i \in I(\Gamma)\cup E(\Gamma) | i \textrm{
  incident on } v\}$ be the set of edges (internal and external)
incident on the vertex $v$. Applying conservation of momentum gives
\bes \langle A_\Gamma(z)f, h \rangle = \frac{1}{\Gamma(z)^{|I(\Gamma)|}}
\int_{M^V }f h \int_0^\infty \sum_{k_1 \ldots k_I = 0}^\infty
\prod_{i=1}^{|I(\Gamma)|}e^{-t_i\lambda_{k_i}} t_i^z dt_i  \\ \prod_{v\in V(\Gamma)}\sum_{j \in f_v; k_j = 0}^\infty a(\{k_j|j \in f_v\})
\dvol(x_1, \ldots, x_{|V(\Gamma)|})\; .\ees Working out the $a(\{k_j|j \in
f_v\})$s re-indexes the eigenvalues in terms of the graphs loop
number, $L$ and loop indices, $l_i$. From here, one can apply the
Schwinger trick, and carry out calculations in a manner similar to
\cite{IZ}, Chapter 6. The operator $A_\Gamma(z)$ is a convolution
product of the $\Delta_M^{-1}$, twisted by the quantities $a(\{k_j|j
\in f_v\})$. Since the trace of $\Delta_M^{-1-z}$ and $a(\{k_j|j \in
f_v\})$ depend only on the metric of $M$, so does $\langle
A_\Gamma(z)f, h\rangle$.
\end{proof}

\begin{remark}
The functions $a(i,j,k)$ are implicitly functions of the metric,
$g(x)$, on $M$. Therefore the regularized operators associated to
Feynman integrals depend on $g(x)$. If the metric is constant, the
regularized operator is independent of the position over $M$. In the
special case where $M$ is a flat manifold, then \bes a(i,\,j,\,k)=
\begin{cases}
1 & \text{if $\sqrt{\lambda_{\sigma(i)}}+\sqrt{\lambda_{\sigma(j)}}=\sqrt{\lambda_{\sigma(k)}}$ for $\sigma \in S_3$} \\ 0
&\text{else.}
\end{cases} \ees
If $M$ is flat, then $\lambda_i= p^2$ is the square of the momentum,
and $a(i,\,j,\,k)$ imposes conservation of momentum at each vertex.
\end{remark}

\begin{corollary}
There is a graph polynomial associated to each Feynman graph on
$M$. The terms of the polynomial differ from the graph polynomials
over flat space-time found in \cite{IZ}, chapter 6, only by the
coefficients, which are functions of $a(i,\, j,\, k)$.\end{corollary}

\subsection{The renormalization bundle}

It remains to construct the renormalization bundle over $M$. The
global divergence structure of the Feynman diagrams is inherited from
the divergence structure of Feynman integrals calculated
locally. Therefore the Hopf algebra of 1PI Feynman diagrams on $M$ is
the same as the Hopf algebra of 1PI Feynman diagrams on $\R^6$.

\begin{remark}
Corollary \ref{opregfr} shows that the operator regularized Feynman
rules assign an operator, $A_\Gamma$ to each $\Gamma \in \h$. This is a
linear map from $S(E) \rightarrow \A$. As shown in \cite{CMbook}, the
\emph{full} Hopf algebra associated to a regularized QFT is
$\tilde{\h} = S(\mathcal{D}'(M))$, is the symmetric algebra on
distributions on $M$. This larger Hopf algebra can be related to $\h$,
and therefore I continue to work with $\h$. \end{remark}

Let $K_M$ be a trivial $G(\A)$ bundle over $M$. If $\gamma(x)$ is a
section, write $\gamma(x) = \varphi_x(z)$ with $\varphi_x(z) \in
G(\A)$ for each $x$. Define $\Delta^*_M \simeq \Delta^* \times M$.
Then $K_M$ can be written as a $G$ principal bundle over $\Delta^*_M$,
with sections $\varphi(x, z) = \varphi_x(z)$.

\begin{lemma}
Let $\gamma(x)$ be a section of $K_M \rightarrow M$. It can be
decomposed into two sections \bes \gamma(x) = \gamma(x)_-^{-1}
\star \gamma(x)_+\;,\ees such that the sections $\gamma(x)_-$
and $\gamma(x)_+$ correspond to the counterterms and renormalized
sections computed locally. \label{MBirkhoff}
\end{lemma}

\begin{proof}
The counterterm and renormalized sections are defined by Birkhoff
decomposition on the fibers \bes \gamma(x)_- = \varphi_x(z)_- \quad
;\quad \gamma(x)_+ = \varphi_x(z)_+ \;. \ees Let $\psi(x)$ be a
section of $K_M \rightarrow M$ defined on $U \subset M$ a coordinate
patch on $U$ with coordinate map $\phi: U \rightarrow \R^6$. Then
$\psi \circ \phi^{-1}$ is a section of $K \rightarrow \Delta^*$. The
decomposition of $\gamma(x)$ can be written $\psi(x)_- = (\psi(x)
\circ \phi^{-1}(x))_-$ and $\psi_+(x)= (\psi(x) \circ
\phi^{-1}(x))_+$.
\end{proof}

Incorporating the renormalization group action, which is uniform over
all of $M$, one has the bundle $P_M \rightarrow B_M$, where $P_M \simeq
\Delta^*_M \times \C^\times \times G$, and $B_M \simeq \Delta^*_M
\times \C^\times$. The renormalization group action on this bundle is, as in
the flat case, \bes t \circ (\varphi(x, z), u) = (t^{Y}\varphi(x,
z), tu) \;.\ees Also, one can write $P_M\rightarrow B_M$ as a
$\tilde{G}(\A)$ bundle over $B_M$. By construction, is a $\C^\times$
equivariant bundle.

\begin{definition}
A section of $K_M \rightarrow M$ has local counterterms if \bes
\frac{\partial}{\partial t} (t^Y \gamma(x))_- = 0\;. \ees
\end{definition}

Notice that $\gamma(x)$ has local counterterms if and only if it has
local counterterms on all fibers, that is, if $\varphi_x(z)$ has local
counterterms for every $x\in M$.

\begin{theorem}
Let $\gamma_{or}(x)$ correspond to the operator regularized
scalar QFT defined globally on $M$. It has local
counterterms. \label{orlocal} \end{theorem}

\begin{proof}
Let $\{U_i, \phi_i(x)\}$ be an atlas of $M$. The atlas defines a
section of $K \rightarrow \Delta^*$ \bes \gamma_{or}(x)|_{U_i}\circ
\phi_i^{-1}(x) = \varphi_{or, U_i}(z) \;, \ees that corresponds to
coordinate patch calculations of operator regularization. Each
$\varphi_{or, U_i}(z)$ has counterterms by Theorem \ref{localthm}. By
Lemma \ref{MBirkhoff}, $\gamma_{or}(x)$ has local counterterms.
\end{proof}

Define a connection on $B_M$ \bes \varphi^*\omega(x, z, t) =
(\varphi(x, z), t)^{-1} \star d (\varphi (x, z), t) \;.\ees Since
$P_M$ is a $\C^\times$ equivariant bundle, I can restrict to the
sections $(t^Y\varphi(x, z), 1)$. Then \bes (t^Y\varphi)^*\omega(x, z) &=&
(t^Y\varphi(x, z))^{\star -1} \star d (t^Y\varphi(x, z)) \in
\Omega^1(TM \oplus \tilde\g(\A)) \\ &=& (t^Y\varphi(x, z))^{\star -1}
\star \left( \sum_{i=1}^6\nabla_i (t^Y\varphi(x, z))dx_i +
\frac{d}{dz} (t^Y\varphi(x, z)) dz + t\frac{d}{dt} (t^Y\varphi(x, z))
dt \right)\;.\ees Unlike the connections defined in \cite{CMbook},
this connection is not flat in general.

The renormalization group flow of a section of this bundle is the
vector field generating the action of the renormalization group on
this bundle.  The renormalization group action on a section
$(\varphi(x,z), s)$ defines a one parameter path in $P_M$, $C_\varphi
= \{(t^Y \varphi(x,z), st)| t \in \C^\times\}$. After restricting to
sections of the form $(t^Y\varphi(x, z), 1)$, the generator of the
renormalization group flow, $\omega_t(\varphi(x,z))$ is the vector
field generated by the logarithmic derivative of the curve
$C_\varphi$.  Write \bes \omega_t(t^Y\varphi(x, z)) =
[(t^Y\varphi(x,z))^{\star -1}\star t \frac{d}{dt}
  (t^Y\varphi(z,x))]|_{t=1} \; .\ees This is exactly the coefficient
of $dt$ in the expression for $(t^Y\varphi)^*\omega(x, z)$.

This vector field is defined for all sections of $P_M \rightarrow
B_M$, but these sections need not correspond to regularization schemes
that have well defined counterterms.

\begin{definition}
The $\beta$-function of a section $\varphi(x,z)$ is \bes
\beta(\varphi(x,z)) = \lim_{z \rightarrow 0} z \,\omega_t(t^Y\varphi(x,
z)) = \lim_{z \rightarrow 0} z [(t^Y\varphi(x,z))^{\star -1}\star t
  \frac{\partial}{\partial t} (t^Y\varphi(z,x))]|_{t=1} \; .\ees
\end{definition}

\begin{theorem}
If $\varphi(x,z)$ has local counterterms, then $\beta(\varphi(x,z))$ is
  well defined. \end{theorem}

\begin{proof}
If $\varphi(x,z)$ has local counterterms, then for each $x\in M$,
$\varphi_x(z)$ has local counterterms. Therefore $\beta(\varphi_x(z))$
is defined, defining $\beta(\varphi(x,z))$ on the fibres.
\end{proof}

In theorem \ref{orlocal}, I show that $\gamma_{or}(x): M \rightarrow
K_M$ has local counterterms. Let $\varphi_{or}(x, z): \Delta_M
\rightarrow K_M$ be the corresponding section of $K_M \rightarrow
\Delta^*_M$. This has local counterterms. Therefore,
$\beta(\varphi_{or, M}(x, z))$ is well defined. It is a vector field
that defines the renormalization group flow over all of $M$.

Let $U \subset M$ be a coordinate patch on $M$ with coordinate map
$\phi: U \rightarrow \R^6$. For another example of a section of $K_M
\rightarrow \Delta^*_M$ with local counterterms, consider \bes
\gamma_{dr, U}(x) : U \rightarrow K_M ,\ees the section of $K_M
\rightarrow M$ corresponding to dimensional regularization on
$U$. Then $\gamma_{dr, U}(x)_-$ has local counterterms, since \bes
\gamma_{dr, U}(x)_- = (\gamma_{dr, U}(x)\circ \phi^{-1}(x))_-\;, \ees
and the righthand side has local counterterms. Let $\varphi_{dr, U}(x,
z): \Delta^* \times U \rightarrow K_M$ be the section corresponding to
$\gamma_{dr, U}(x)$. The generator of the renormalization group flow
for dimensional regularization on $U$, $\beta(\varphi_{dr, U}(x, z))$
is well defined. However this flow cannot be extended globally over
$M$.

\begin{remark}
Because of the difference in definition between the physical and
geometric $\beta$-functions, $\beta(\varphi_{or, M}(x, z))$ does not
resolve the problems that arise in calculating the physical
$\beta$-function globally. In fact, when $M \neq \R^6$, $\beta(\varphi_{or, M}(x, z))$ does not correspond to the scale dependence of the coupling constant. However, it is the vector field the governs the \emph{Lagrangian's} scale dependence \cite{dynkin}.
\end{remark}

\section{Conformal changes to the metric \label{conformal}}

In this section, I extend the above analysis to operator
regularization of a conformally coupled field theory over a background manifold
$M$, where the metric is unspecified. Unlike in the previous section, the renormalization mass
parameter is not constant over $M$. This is done by allowing
conformal changes to the metric of the manifold. There has been a lot
of work done studying conformal Lagrangians, and the associated
conformal anomalies. For instance, see \cite{Pan}, \cite{ParRos}, and
\cite{BranOr}. The Laplacian $\Delta_M$ is no longer sufficient to
properly determine the dynamics of a conformally coupled field theory. I define
a suitable conformally corrected Laplacian, a differential operator
under which the free Lagrangian for the scalar field theory is
conformally invariant. The renormalization group flow for a conformally coupled
scalar field theory can be geometrically encoded for the theory thus
defined.

In the previous section, the regularized Lagrangian had the form \be
L(z) = \int_M \left[\frac{1}{2} \phi(x)( \Delta)^{1+z}_M \phi(x)
  +\lambda\mu^{-2z} \phi(x)^3 \right] \mu^{2z}\dvol(x) \; ,\ee
as in equation \eqref{scalingor}. The factor of $\mu^{2z}$
corresponds to a scaling of the metric by the constant
$\mu^{\frac{2z}{3}}$. Instead, scale the metric by a conformal
factor $e^{2f(x)}$, where $f \in C^\infty(M)$. This changes the
renormalization bundle of the previous section, as the renormalization
mass parameter $\C^\times$ no longer sits trivially over $M$.  To
understand this new bundle, I introduce the language of densities over
the manifold $M$, and write the renormalization bundle for conformally coupled
field theories in the language of density bundles over $M$.

\subsection{Densities}
In this section, I review some properties of densities. While
compactness and orientability are not necessary for the arguments of
the following sections, I maintain the conventions of the previous
sections and let $M$ be a smooth, compact, oriented Riemannian
$n$-manifold. Let $\fr(M)$ be the frame bundle over $M$. It is a
$\Gl_n(\R)$ principal bundle over $M$.

\begin{definition} For an orientable manifold and for any $r \in \R$ \bes |\det|^{r/n} : \Gl_n(\R) \to \R_+^\times \ees defines a line
bundle, $\R(r)$, or $r$-densities over $M$. \end{definition}

The bundle $\R(r)$ can be trivialized by choosing a metric,
$g$, for $M$. Let $ \phi$ be a section of $\R(r) \rightarrow M$. Given
a choice of $g$, it can be written uniquely as \bes \phi =
f|g|^\frac{r}{2n}\ees for some $f \in C^\infty(M)$.

If $\phi$ is a continuous section of $\R(r)$, then for any $s > 0, \;
|\phi|^s$ is a continuous section of $\R(rs)$. These sections can be
given a Banach norm. For $n \geq r > 0$, \bes ||\phi ||_{n/r} :=
(\int_M |\phi|^{n/r})^{r/n} \;\ees This becomes apparent under a
trivialization \bes ||\phi||_{n/r} = (\int_M
(|f||g|^\frac{r}{2n})^{n/r})^{r/n} = (\int |f|^{n/r} \dvol(g))^{r/n}
\; .\ees When $r=0$, the norm is given by the classical essential
supremum.

\begin{definition}
I write $\bL(r)$ for the Lebesgue space of $r$ densities, with these
norms \bes \bL(r) = \overline{(\phi \in \R(r) ; ||\phi||_{n/r} <
  \infty)} \;.\ees
\end{definition}

 In this terminology, $n$-forms become $n$-densities, the Banach space
 dual of $\bL(d)$ is $\bL(n-d)$. Sections of $\R(\frac{n}{2})$ define
 a Hilbert space $\bL(\frac{n}{2})$ with inner product \bes \langle
 \phi, \psi \rangle = \int_M \phi \psi \; .\ees This inner product is
 independent of the Riemannian metric. A choice of $g$ defines an
 isometry with the classical Lebesgue space $L^2(M,g)$. Let
 $\phi=f|g|^{\frac{1}{2}}$ and $\psi = h|g|^{\frac{1}{2}}$. The inner
 product is \bes \langle \phi, \psi \rangle_g = \int_M f
 |g|^\frac{1}{4} h |g|^\frac{1}{4}dx_1 \wedge \ldots \wedge dx_n \;
 .\ees

Finally, there is a linear operator \be \phi
\mapsto |g|^\frac{d_1 - d_0}{2n} \phi \label{densitychange} \ee that maps
smooth sections of density $d_0$ to those of density $d_1$.  When $d_1
\geq d_0$ it defines a continuous linear map from $\bL(d_0)$ to
$\bL(d_1)$.

\subsection{Effect of conformal changes on the Lagrangian}

I can use this formalism to study how the Lagrangian varies under
conformal changes to the metric \bes g \rightarrow e^{f(x)}g \quad ;
\quad f(x) \in C^\infty(M) \;. \ees For ease of notation, let $u =
e^f$. The Lagrangian density for renormalizable scalar field theory on
an $n$-dimensional Riemannian metric is given by \bes \mathcal{L}_M(g) =
\frac{1}{2}\phi(x)(-\Delta(g) + m^2(x) )\phi(x) + \lambda(x)
\phi^\frac{2n}{n-2}(x) \;, \ees where $\phi(x)$ is a $\frac{n-2}{n}$
density, $m(x)$ a 1 density, and $\lambda(x)$ a 0 density on $M$, and
$\Delta(g)$ is the Laplacian on $M$ with respect to the metric $g$.
As before, I write $\Delta_M(g) = -\Delta(g) + m^2(x)$. The density
$\phi$ is raised to an integral power only when $n \in \{3, \, 4,\,
6\}$.

Yamabe \cite{Ya} constructs a conformally coupled free Lagrangian
density \be \mathcal{L}_M(g) = \phi \; [\Delta_M(g) - \frac{1}{4}
  \frac{n-2}{n-1} R(g)] \; \phi  \label{yamabe} \ee for $\phi \in
C^\infty(M)$, and $R(g)$ the scalar curvature on $M$. This is invariant under conformal rescaling $g \mapsto
\bg = e^{2f(x)}g, \; \phi \mapsto \bF = e^{\frac{n-2}{2}f}\phi$, where
$f \in C^\infty(M)$. This is a standard form for weak conformal coupling, \cite{HollandsWald03}

\begin{definition}
Write the conformally coupled Laplacian \bes \Delta_{[g]} =
\Delta_M(g) - \frac{1}{4}\frac{n-2}{n-1}R(g)\; . \ees The $[g]$
subscript indicates that the Laplace operator depends only on the
conformal equivalence class of $g$.  \end{definition}

The operator \bes \Delta_{[g]} \; : \; \bL(\frac{n-2}{2})
\rightarrow \bL (\frac{n+2}{2}) \ees it is a quadratic form on
$\bL(\frac{n-2}{2})$. This is problematic. In order to carry out the
arguments from section \ref{regcomp}, $\Delta_{[g]}$ must be a
self-adjoint operator acting on the Hilbert space
$\bL(\frac{n}{2})$. Fortunately, such an operator can be built out of
$\Delta_{[g]}$.

\begin{theorem}
The operator $Y_g = |g|^{-\frac{1}{2n}} (\Delta_{[g]})
|g|^{-\frac{1}{2n}}$ is a self adjoint operator on $\bL(\frac{n}{2})$.
\end{theorem}

\begin{proof}
By equation \eqref{densitychange}, \bes |g|^{\frac{1}{2n}}\phi \in
\bL(\frac{n}{2})\; .\ees Rewrite the free Lagrangian density in
\eqref{yamabe} as \bes \mathcal{L}_M(g) = \phi
|g|^{\frac{1}{2n}}|g|^{\frac{-1}{2n}}(\Delta_{[g]}
-m^2)|g|^{\frac{-1}{2n}}|g|^{\frac{1}{2n}}\phi \;. \ees Now I can
define an operator \bes Y_g := |g|^{-\frac{1}{2n}} (\Delta_{[g]}-m^2)
|g|^{-\frac{1}{2n}} \ees that acts on the Hilbert space
$\bL(\frac{n}{2})$.
\end{proof}

The operator $Y_g$ can be raised to a complex power. As before,
$\rm{Tr} \, Y_g^{1+z}$ has simple poles in $z$. Following the same
arguments as in Corollary \ref{distregfr}, $\langle f, Y^{1+z}_g
g\rangle \in \C\{\{z\}\}$, or that the corresponding kernel has poles
at $z= 0$.  However, since $\phi \in \bL(\frac{n-2}{2})$, the
self-adjoint operator in the Lagrangian must be a quadratic form on
$\bL(\frac{n-2}{2})$.

\begin{definition}
The operator \bes \tilde{Y}_g(z) = |g|^\frac{1}{2n} Y_g^{1+z}
|g|^{\frac{1}{2n}} \; , \ees is a self adjoint operator on
$\bL(\frac{n-2}{2})$.
\end{definition}

Under the conformal change of metric \bes g \mapsto u^2 g=\bg \ees
$Y_g$ transforms as \bes Y_g \rightarrow u^{-1}Y_g u^{-1} = Y_\bg \ees
Then $\tilde{Y}_\bg(z)$ is, \be \tilde{Y}_\bg(z) =
|g|^\frac{1}{2n}u\left(u^{-1}Y_gu^{-1} \right)^{1+z} u|g|^\frac{1}{2n}
\;\label{nonconflag} .\ee  The
expression $\phi \tilde{Y}_g(z)\phi$ is now a $n+2z$ density. Thus
it behaves nicely under operator regularization.

The kernel of this operator is defined by a
family of pseudo-differential operators with top symbol \bes \xi
\mapsto |\xi|_g^{2+2z} \;.  \ees

The appropriate Laplacian for a conformally coupled scalar field theory is
$\tilde{Y}_g$.

\begin{theorem}
For a general $u=e^f$, let $\bg = ug$. Then $\tilde{Y}_\bg$ can be
expanded as a Taylor series in $f$ as \bes \tilde{Y}_\bg(f,z) =
e^{-2fz}\tilde{Y}_g(z) \;.\ees \end{theorem}

\begin{proof}
Recall that $u = e^{f(x)}$. Calculate the terms of the Taylor series
of $\tilde{Y}_\bg(z)$ at $f=0$.

The $0^{th}$ order term is given by evaluating $\tilde{Y}_\bg(z)$ at
$f =0$. This gives \bes |g|^\frac{1}{2n} Y_g^{1+z}|g|^\frac{1}{2n} =
\tilde{Y}_g(z)\ees

The $n^{th}$ derivative of $\tilde{Y}_\bg(z)$ with respect to $f$,
evaluated at $f=0$ is \bes 2^n(1-(1+ z))^n\tilde{Y}_g(z) = 2^n(-z)^n
\tilde{Y}_g(z)\; .\ees Writing this out as a Taylor expansion

\bes \tilde{Y}_\bg(z)= \sum_{n=0}^\infty \frac{(-2zf)^n}{n!}
\tilde{Y}_g(z) = (\sum_{n=0}^\infty \frac{(-2zf)^n}{n!})
\tilde{Y}_g(z) = e^{-2fz} \tilde{Y}_g(z) \; .\ees
\end{proof}

Now I can define the operator regularized Lagrangian density.

\begin{theorem}
The operator regularized Lagrangian density for a conformally coupled field
theory is \bes \mathcal{L}_M(z,g) = \phi
\tilde{Y}_g(z) \phi + \lambda(x) \phi^{\frac{2n}{n-2}}\;.\ees \end{theorem}

\begin{proof}
I have shown that $\tilde{Y}_g(z)$ is a self adjoint operator on
$\bL(\frac{n-2}{2})$.  The Lagrangian \be L_M(z, [g]) = \int_Mu^{-2z}
\left[\phi \tilde{Y}_g(z) \phi + u^{2z} \lambda \phi^\frac{2n}{n-2}
  \right] \;, \label{disp1}\ee is invariant under a conformal change
of metric.
\end{proof}

\subsection{The renormalization group flow}

Finally, I construct a renormalization bundle that encodes the
renormalization group flow of a scalar conformally coupled field theory under
the operator regularization described above. Let $\varphi_{or, [M]}(x,
z)$ correspond to the operator regularized Feynman rules on $M$.

\begin{definition}
Let $\mathcal{K}_M \simeq G(\A) \times_{GL_n(\R)} \fr(M) \rightarrow
M$ be a bundle over $M$. Choosing a metric on $M$ defines a bundle
isomorphism to $K_M \rightarrow M$.
\end{definition}

\begin{theorem}This can be written as a section of the bundle
$\mathcal{K}_M \rightarrow \Delta^*_M$.  \end{theorem}

\begin{proof}
Let $\varphi_g(x)$ be a section of $K_M \rightarrow M$, where $M$
has the metric $g$. There is a representation $\rho$ of $GL_n(\C)$ on
$G(\A)$ by the action \bes \rho: GL_n(\C) \times G(\A) &\rightarrow&
G(\A) \\ (\nu , \varphi_{M,g}(x,z)) &\rightarrow & \varphi_{M,g'}(x,z)
\;,\ees where $g' = \nu^{-1}g(x)\nu$. The map $\varphi_g'(x,z)$ is
also a section of $K_M \rightarrow M$. The representation $\rho$
defines a vector representation \bes \mathcal{K}_M\rightarrow M\; \ees
that captures the $GL_n(\C)$ action on the fibers of $K_{M, x}$ over
$x \in M$, when the metric on $M$ is not predetermined.  If $g'(x) =
\nu(x)^{-1}g(x) \nu(x)$, then both $\varphi_g(x,z)$ and
$\varphi_{g'}(x,z)$ are sections of $\mathcal{K_M}\rightarrow M$.
\end{proof}

Since the background manifold for a conformally coupled field theory is not
given a fixed metric, this is the appropriate renormalization bundle
for this case.

\begin{theorem}
A section $\varphi_g(x, z)$ of $\mathcal{K}_M \rightarrow M$ can be
Birkhoff decomposed $\varphi_g(x, z)_-$ and $\varphi_g(x, z)_+ $such that
the decompositions agree on the fibers over $M$.\end{theorem}
\begin{proof}
The fibres of $\mathcal{K}_{M, x} \simeq G(\A)$, so Birkhoff
decomposition is well defined. The action of $G(\A)$ on
$\mathcal{K}_M$ ensures that $\varphi_g(x, z)_-$ and $\varphi_g(x,
z)_+$ are sections.
\end{proof}

The renormalization group is now $C^\infty(M, \C^\times)$, under pointwise
multiplication. The coupling constant $\lambda$ transforms as \bes
\lambda \rightarrow e^{f(x)} \lambda \ees under a conformal change of
metric.

\begin{definition}
Let $\mathcal{B}_M \simeq \C^\times(1) \times_{GL_n(\C)} \Delta^*_M$.
\end{definition}

The action of the renormalization group on $\mathcal{K}_M$ gives the
bundle \bes \mathcal{P}_M \simeq \mathcal{K}_M \times \C^\times(1)
\rightarrow \mathcal{B}_M \;.  \ees The sections of this bundle are
$(\varphi_g(x, z), e^{f(x)})$. Now the scale factor $u = e^{f(x)}$ is
a function of $x$, one cannot factor out the renormalization mass as
before.  The action of the renormalization group is \bes u(x) \circ
(\varphi_g(x, z), e^{f(x)}) = (\varphi_{ug}(x, z), ue^{f(x)}) \;.\ees
These are no longer equivariant under the action of
$\C^\times(1)$. This also causes problems with the counterterms for
this theory.

For a renormalization scheme on a regularized theory to be physically
meaningful, as before, the dimension of the counterterms should not
depend on the regularization parameter, and the renormalized values
should be finite.

\begin{definition}
A conformally coupled field theory has local counterterms if \bes
\frac{\partial}{\partial u(x)} (\varphi_{ug}(x,z))_- = 0 \;. \ees
\end{definition}

The definition of the projector $\pi: \A \rightarrow \A_-$ means that
the renormalized part of the regulated theory under BPHZ is
finite. However, conformal operator regularization does not have local
counterterms under BPHZ.  Nonetheless, the renormalization group flow
of $\varphi$ on $\mathcal{P}_M\rightarrow \mathcal{B_M}$ can be
calculated by the logarithmic differential as \bes \left(
(\varphi_g(x, z), e^{f(x)})^{\star -1} \star d (\varphi_g(x, z),
e^{f(x)}) \right) |_{f=0} \;. \ees

\bibliographystyle{amsplain}
\bibliography{Bibliography}{}

\end{document}